\documentclass{pas}

\usepackage{tikz}
\usetikzlibrary{shapes.geometric, arrows}
\tikzstyle{block} = [rectangle, draw,
    text width=5em, text centered, minimum height=4em]
\tikzstyle{line} = [draw, -latex']
\tikzstyle{question}=[draw, shape=regular polygon, regular polygon sides=3,draw,thick,inner sep=0pt,minimum
size=8cm]

\usepackage{multirow}
\usepackage{amsmath}
\usepackage{natbib}
\usepackage[T1]{fontenc}
\begin{document}

\lefttitle{Publications of the Astronomical Society of Australia}
\righttitle{D. Giancono et al}

\jnlPage{1}{13}
\jnlDoiYr{2026}
\doival{10.1017/pasa.2026.10162}


\title{Photometry of Fireballs using High Frame Rate Cameras}

\author{
\sn{D. } \gn{Giancono}$^{1}$ 
\sn{H. } \gn{Devillepoix}$^{2}$
\sn{R} \gn{Howie}$^{3}$
\sn{D} \gn{Vida}$^{4}$
\sn{D} \gn{Rollinson}$^{5}$
}

\affil{
$^1$Curtin Institute of Radio Astronomy/Space Science and Technology Centre, Curtin University, 6845, WA , Australia \\
$^2$Curtin Institute of Radio Astronomy/Space Science and Technology Centre, Curtin University, 6845, WA , Australia \\
$^3$Space Science and Technology Centre, Curtin University, 6845, WA , Australia \\
$^4$Department of Physics and Astronomy, University of Western Ontario, N6A 3K7, Ontario, Canada \\
$^5$Perth Observatory Volunteer Group, Bickley, Western Australia, Australia
}

\corresp{D. Giancono, Email: d.giancono@gmail.com}



\begin{abstract}
Fast sampling photometry is a key observable for characterising fireballs, particularly their fragmentation episodes, which are strongly connected to the internal structure of the meteoroid and its physical properties.
Accurate photometric measurements remain a challenge due to the large dynamic range required (upwards of 10 stellar magnitudes), driving operational complexity and cost.
We have developed a system using an all-sky camera operating at up to 500 frames per second, featuring a novel implementation of Detection Localised Auto-brightness Control. The large data throughput is managed by custom software that performs transient detection, region-of-interest saving, and real-time photometry. We present results from two field deployments: the first validates the system's photometric accuracy against conventional 30 frames per second cameras, while the second demonstrates the successful implementation of Detection Localised Auto-brightness Control in capturing a bright, magnitude -15 fireball with minimal saturation. With the Detection Localised Auto-brightness Control, the system achieves an effective dynamic range between apparent magnitudes of approximately -3 to -17, allowing it to capture light curves with minimal saturation for most fireballs, excluding rare superbolides. The resulting high-quality light curve enabled a successful semi-empirical fragmentation analysis verifying the system's ability to provide data for detailed physical modelling. The primary application for this validated system will be as a core component of the Global Fireball Observatory's next-generation instrumentation. The intention is to deploy it in a hybrid observatory, operating alongside a dedicated high-resolution astrometric camera. This configuration will allow the network to simultaneously capture precise trajectory data for orbit and fall-line calculations and acquire complete, unsaturated high dynamic range light curves at high temporal resolution for detailed physical analysis, combining the strengths of both systems.
\end{abstract}


\maketitle

\section{Introduction}
Understanding the physical structure and fragmentation behaviour of meteoroids during atmospheric entry is critical for characterising small Solar System bodies and ultimately constraining their composition and source regions. 
Photometric fireball and meteor observations provide a direct means of probing meteoroid ablation processes, mechanical strength, and fragmentation dynamics, key factors that influence whether a meteorite survives to be recovered and analysed \cite{campbell-brown_meteoroid_2019}. 
Short-term variations in fireball light curves, such as flares and sudden brightness changes, are signatures of fragmentation events and offer insight into the nature of mass loss and grain size distributions. These physical properties, in turn, inform models of meteoroid composition and contribute to accurate orbital reconstruction  \citep{asher_modelling_2019}. 
Complete, unsaturated, high temporal resolution light curves are therefore valuable not only for modelling fragmentation and ablation processes but also for estimating the initial mass, composition, and ultimately, the origin of meteoroids within the Solar System \citep{spurny_zdar_2020}.
Long-exposure imagery and video recordings are typical methods for obtaining photometric data from fireballs \citep{borovicka_data_2022} \citep{colas_fripon_2020}.
These image-based observations also enable the estimation of a fireball's trajectory and velocity throughout its luminous flight by capturing the event from multiple vantage points and applying triangulation to the observed meteor path.
This, in turn,  allows for the calculation of the meteoroid’s heliocentric orbit prior to atmospheric entry, linking observed fireballs to their potential source regions \citep{howie_submillisecond_2017}.
The fall position of any surviving mass can be estimated by dark flight modelling, which involves the calculation of the fireball's final mass, as well as wind modelling, which can help with the efficient recovery of meteorite samples with a known parent body \citep{towner_dark-flight_2022}. 
Estimating the initial mass of the meteoroid, crucial for both orbital and fragmentation analysis, can be achieved using integrated brightness during bright flight (photometric mass), as well as meteor deceleration (dynamic mass) \citep{gritsevich_validity_2008}.
However, dynamic mass estimates often assume a single-body model and can significantly underestimate mass in the presence of fragmentation. Conversely, photometric mass estimates may suffer from saturation effects in long-exposure and video imagery, leading to further uncertainties \citep{borovickaInstrumentallyRecordedFall2015}.
Meteor fragmentation is responsible for sudden changes in mass and often appears as short flaring events or abrupt decelerations during bright flight. These events not only allow for strength estimation based on dynamic pressure but also reveal internal structure regardless of whether any surviving mass is recovered \citep{borovicka_two_2020}.
Resolving these rapid events remains a challenge due to the poor temporal resolution of commonly used imaging techniques \citep{spurny_bunburra_2012}.
Occulted long-exposure imagery uses mechanical or electro-optical shutters to interrupt camera exposure periodically, embedding timing information into the streak of a meteor trail as it moves across the sky \citep{howie_submillisecond_2017}. 
The Desert Fireball Network (DFN) and the subsequent Global Fireball Observatory (GFO) are primary examples of networks based on this principle; this successful, low-cost, global system was engineered around a timing solution using a liquid crystal shutter to encode a precise de Bruijn timecode directly onto the high-resolution astrometric image. This design choice eliminated the need for complex secondary timing hardware but prioritised astrometry by accepting a key trade-off: it inherently sacrifices the complete photometric record in the shutter gaps. 
The generation of the final photometric record using this method involves three distinct processing stages. First, an instrumental light curve is produced by extracting raw intensity values from the occulted trail, typically by summing pixel counts within each shutter segment and subtracting the background. Second, these values are converted into an apparent magnitude light curve by calibrating the instrumental flux against reference stars identified in either the detection image or separate calibration frames, while applying a correction for atmospheric extinction. In the final step, a range-corrected absolute magnitude (normalized to 100 km) is computed to account for the varying distance of the fireball from the observer, ensuring that light curves recorded from different locations are directly comparable.
Video camera imagery has the benefit of not requiring optical shutters for deriving timing information as each frame within a video file can be processed separately by extracting pixel values from the fireball in each frame. This eliminates the issue of smearing between photometric segments which can make accurate photometry difficult in long-exposure imagery, especially when the fireball becomes bright and its apparent size on the sensor is large. The light curve sampling rate of these methods is usually less than 30Hz, and both methods are prone to saturation during bright fireball events which is usually caused by the limited dynamic range of the sensors (usually 8 or 10 bits). This limits the usefulness of both methods for fragmentation identification and photometric mass calculations during a fireball's entire bright flight \citep{howie_how_2017,colas_fripon_2020,vida_global_2021,spurny_zdar_2020}. 
An alternative approach is to use all-sky radiometers which record the brightness of the entire sky during a fireball event as a time series, and typically use photomultiplier tubes (PMT) due to their large dynamic range and high sensitivity. These sensors operate at high sampling rates of 500 Hz or 5000 Hz, representing far better temporal resolution compared to image-based methods \citep{spurny_zdar_2020}. Calibration is performed by taking absolutely calibrated light curves from image sources and aligning the all-sky brightness light curve to the image-based light curve during periods in which the imagery is unsaturated. While useful for fragmentation identification due to its superior temporal resolution and dynamic range, several drawbacks exist when using the PMT data for generating light curves compared to image-based methods such as requiring protection from daylight and exhibiting different sensitivities across individual tubes \citep{spurny_bunburra_2012}. On nights where large amounts of ambient light are present (usually caused by the presence of the Moon), the sensitivity of the device must be reduced to ensure that the sensor does not saturate \citep{spurny_analysis_2010}. To mitigate some of these operating requirements, photodiodes have also been used as all-sky brightness sensors as they do not require high voltage power supplies, protection from daylight, and are physically more robust than photomultiplier tubes \citep{popowicz_enhancing_2025}. However, they still require sensitivity control circuits, noise cancelling techniques, and often have a spectral response that does not match well to the spectral emission of fireballs. \citep{segon_first_2018,rault_little_2020,vida_low-cost_2015,buchan_developing_2019}. Like PMT-based systems, photodiode sensors must also be calibrated using absolutely calibrated image-based light curves, typically during unsaturated portions of the fireball's bright flight, to produce meaningful photometric data.
Complete, high-cadence light curves are an essential input for fireball entry modelling, notably for semi-empirical ablation and fragmentation models \citep{borovicka_kosice_2013,asher_modelling_2019}.
With the GFO’s native hardware currently not providing this data, detailed modelling is only possible when external instruments fill the photometric data gap. One of these instruments is the space-based Geostationary Lightning Mapper (GLM), a 500 FPS very low-resolution imager designed for lightning detection. The capability of the GLM to detect and characterise bolides was first systematically demonstrated by \cite{jenniskens_detection_2018}, establishing it as a valuable resource for meteor science.
In key events where the ground-based network cameras saturated, this supplementary GLM data provided the necessary unsaturated light curve, enabling accurate estimates of a meteoroid’s physical properties through fragmentation modelling \cite{vida_direct_2023}.
Because the GLM covers only a fraction of the Earth (primarily the Americas) and has a sensitivity limit of approximately magnitude -14, it cannot be relied upon as a general solution for fireball observation.
This illustrates the strategic need to develop an internal, integrated, low-cost hardware solution that replicates the function of a PMT-based radiometer, securing the complete photometric record required for fragmentation studies without the complexity of a hybrid PMT system.
High frame rate imaging, using modern CMOS cameras, presents a pathway to design such an integrated solution.
They provide the high temporal resolution of a radiometer while remaining an imaging system that can be photometrically calibrated against background stars, eliminating the need for inter-instrument calibration.
Furthermore, the primary limitation of digital imagers, limited dynamic range, can be overcome by implementing real-time Detection Localised Auto-brightness Control (DLAC) which can dynamically change exposure and gain settings based on the brightness of the fireball within the aperture of a detected event.
This paper presents a system prototype for the next generation of GFO instrumentation, which has achieved Technology Readiness Level 7 (TRL-7) by being successfully demonstrated in an operational environment, thereby validating its integrated capabilities.
We first demonstrate that this high frame rate system, operating at 470 frames per second (FPS), produces calibrated photometry comparable to existing 30 FPS systems while capturing the high-cadence fluctuations associated with fragmentation.
We then present observations captured by an improved system, that is able to acquire data at 500FPS, and demonstrates the successful implementation of DLAC during a fireball event. 
We show that this technique successfully captures the full light curve with minimal saturation, proving the system can provide all data necessary for detailed fragmentation modelling from a single, robust, and low-cost instrument.
This validated approach solves the primary observational bottleneck of existing shuttered long-exposure camera systems, which saturate during bright events and miss photometric data in occulted gaps.

\section{Instrument Description}

\subsection{Hardware}
\begin{figure*}[ht!]
    \centering
    \includegraphics[width=\textwidth, trim={0 0 0 14cm},clip]{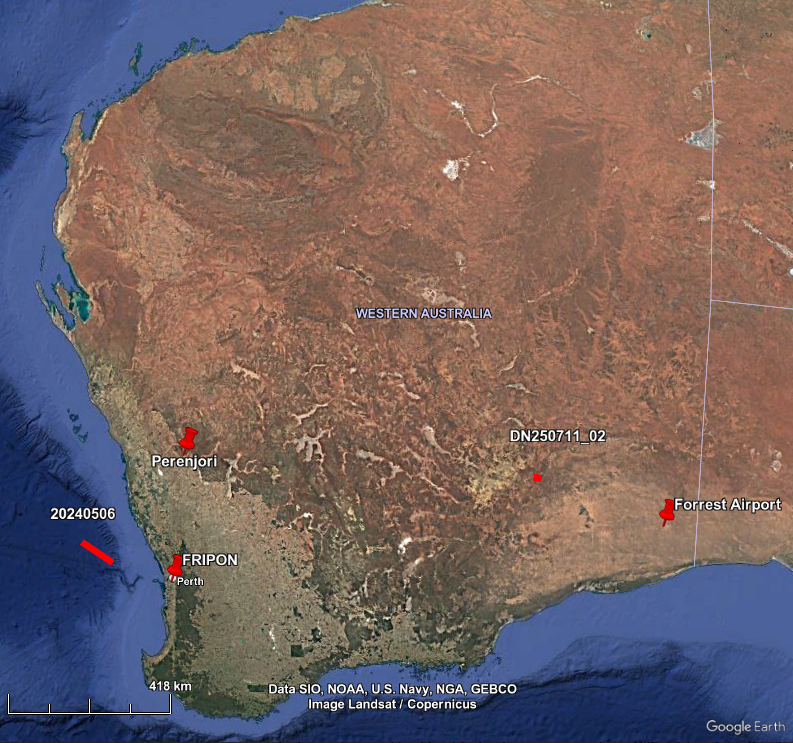}
    \caption{Locations of the Perenjori and Forrest Airport prototypes, as well as the FRIPON station used for photometric verification. The trajectories of 20240506 and DN250711\_02 are also shown.}
    \label{fig:stations}
\end{figure*}

To evaluate the use of high frame rate cameras for imaging fireballs, two prototype instruments were developed using the same off-the-shelf camera but with different optical and operational configurations.
The initial prototype was installed at an existing DFN site near Perenjori, Western Australia as shown in Figure \ref{fig:stations}, and is powered by a solar panel and battery system.
It consists of an Allied Vision Alvium U-052 USB machine vision camera (Sony IMX426 CMOS global shutter sensor), the characteristics of which are outlined in Table \ref{tab:camera_specs_basic}. This camera was selected due to its USB interface, high frame rate capability, high saturation capacity, 12 bit depth output, global shutter mode, and relative affordability. 
\begin{table}
\centering
\caption{Allied Vision Alvium 1800 U-052 Camera specifications and imaging performance}
\label{tab:camera_specs_basic}
\begin{tabular}{|l|l|}
\hline
\textbf{Specification} & \textbf{Value} \\
\hline
Camera Model              & Alvium 1800 U-052 \\
Resolution                & 816 (H) × 624 (V) \\
Frame Rate                & 688 fps at full resolution in Mono8\\
Sensor Type               & CMOS \\
Sensor Name               & Sony IMX426 \\
Sensor Size              & Type 1/1.7 \\
Shutter Mode              & GS (Global shutter) \\
Pixel Size               & 9 µm × 9 µm \\
Quantum Efficiency        & 73\% at 529 nm \\
Temporal Dark Noise       & 21.8 e\\ 
Saturation Capacity       & 100000 e\\
Absolute Sensitivity Threshold & 23.7 e\\
Bit Depth                 & 8-bit, 10-bit, 12-bit; \\
Monochrome Pixel Formats  & Mono8, Mono10, Mono12 \\
\hline
\end{tabular}
\end{table}
The camera was connected to a Fujinon FE185C046HA 1.4 mm fish-eye lens, giving an all-sky view with a field of view in excess of 180 degrees and a pixel scale of approximately 22 arcmin. To provide a weatherproof seal, the lens is mounted in a custom aluminium flange and bonded to the front glass element with a flexible silicone sealant. It is configured to take 470 FPS, with an exposure period of 1925.475 µs. With this exposure period the camera operates with a 90.5\% duty cycle, with the remaining time dedicated to the readout of frames. It captures images in a 12-bit pixel format and the saved images have an effective bit-depth of 12. The frame rate of 470 FPS was selected as it is the maximum achievable with the chosen camera using a USB 3.0 interface. Based on the operational experience from the initial prototype, a second, more advanced system was developed and deployed to Forrest Airport in the Nullarbor (see Figure \ref{fig:stations}). This iteration incorporated several key improvements, most notably the implementation of the Detection Localised Auto-brightness Control (DLAC) algorithm to address the critical challenge of sensor saturation. Furthermore, to simplify the hardware for easier manufacturing and field deployment, the custom-sealed Fujinon lens was replaced with an IP-rated, waterproof Commonlands CIL293 1.3 mm stereographic fisheye lens, which provides an all-sky field of view without requiring a bespoke weatherproofing assembly. This system uses the same camera but is configured with a region-of-interest (ROI) crop, allowing it to operate at a fixed 500 FPS. The maximum exposure period is 1800 µs, and the duty cycle of the system varies dynamically as the DLAC algorithm decreases the exposure to manage bright events. Both prototypes use an NTP-synchronised local computer clock to provide absolute timing of detections.

\begin{figure}
    \centering
    \includegraphics[trim={14cm 0 15cm 0},clip,width=4cm,height=6cm]{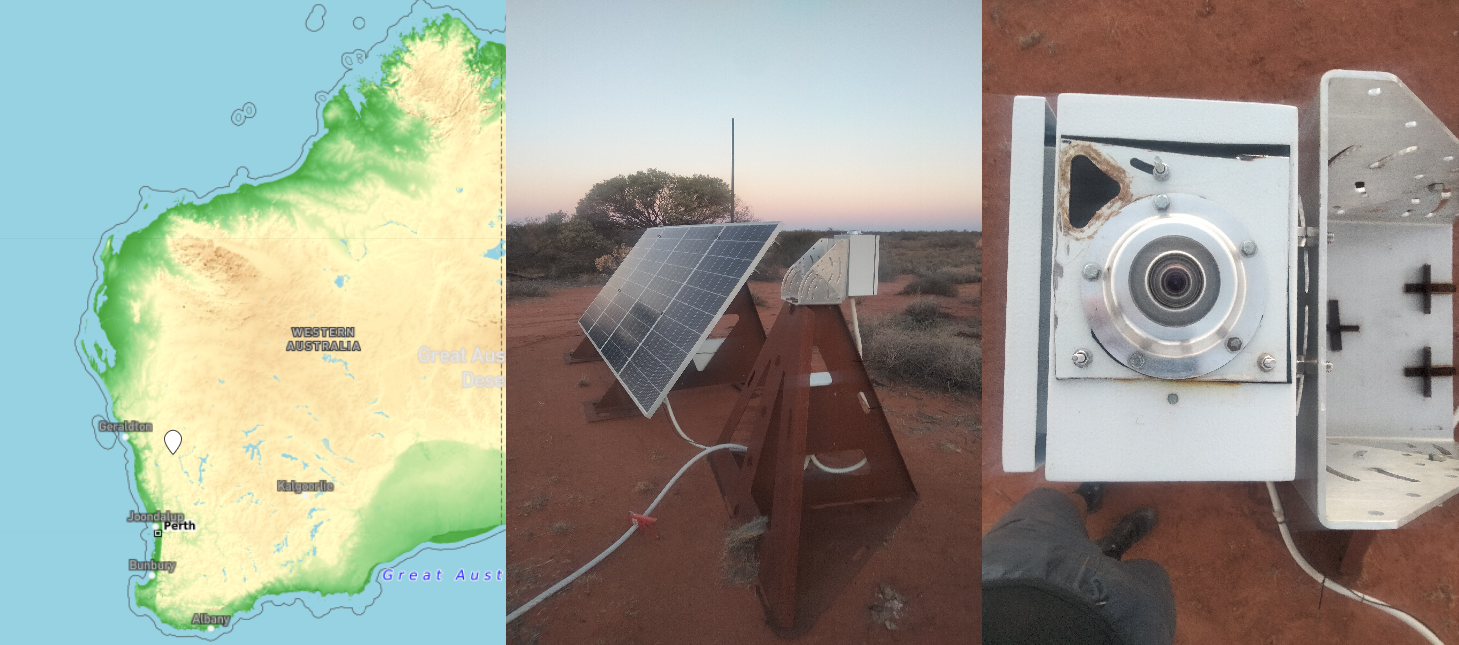}
    \includegraphics[trim={0 0 0 0},clip,width=4cm,height=6cm]{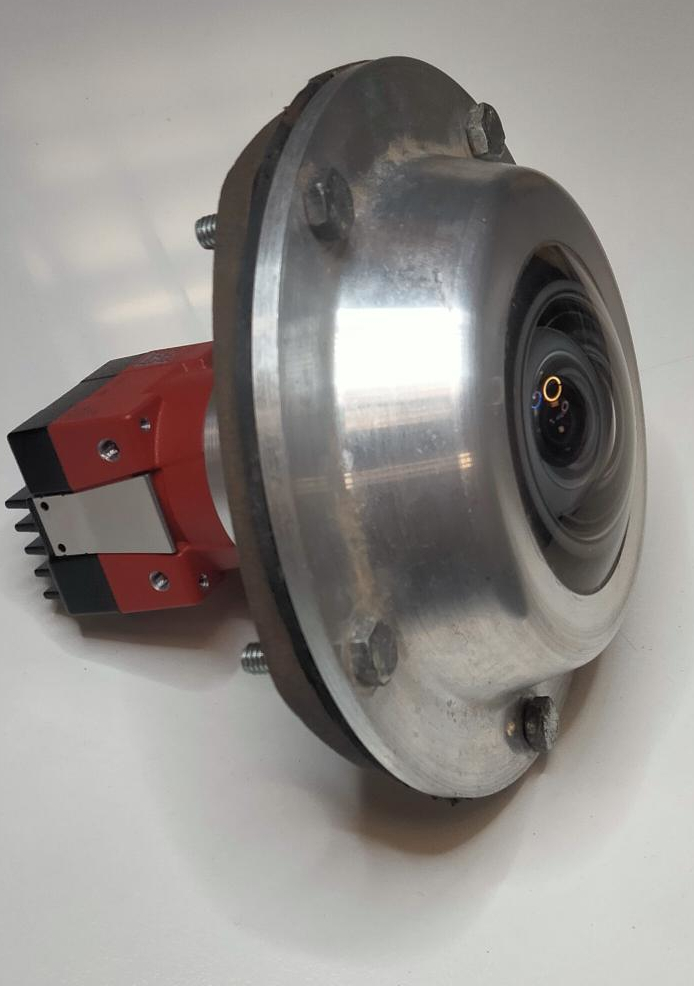}
    \caption{\textbf{Left:} Prototype installation near Perenjori, Western Australia. \textbf{Right:} Allied Vision Alvium U-052 USB machine vision camera connected to a weatherproofed Fujinon FE185C046HA lens assembly.}
    \label{fig:radcam}
\end{figure}

\subsection{Fireball Detection and Processing}
\begin{figure}
    \centering
    \includegraphics[width=\columnwidth]{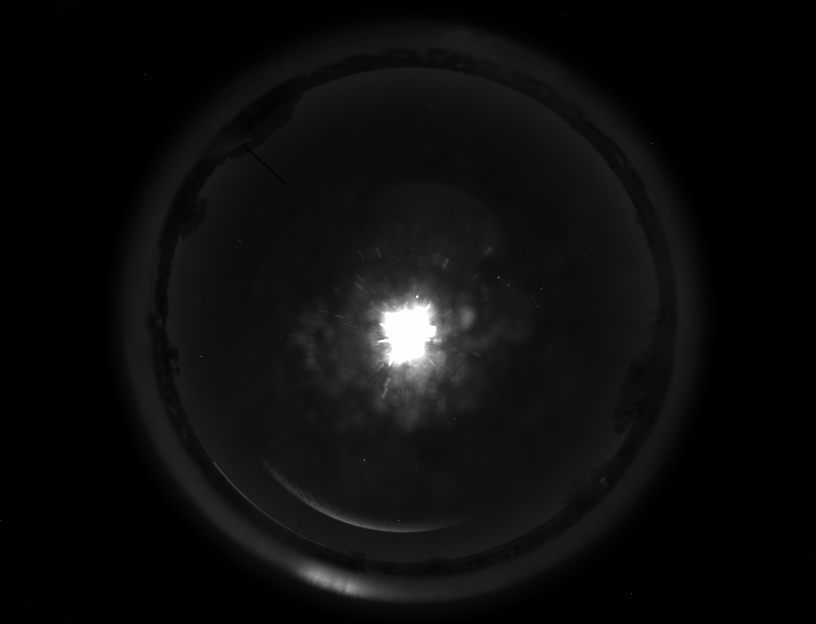} 
    \includegraphics[width=\columnwidth]{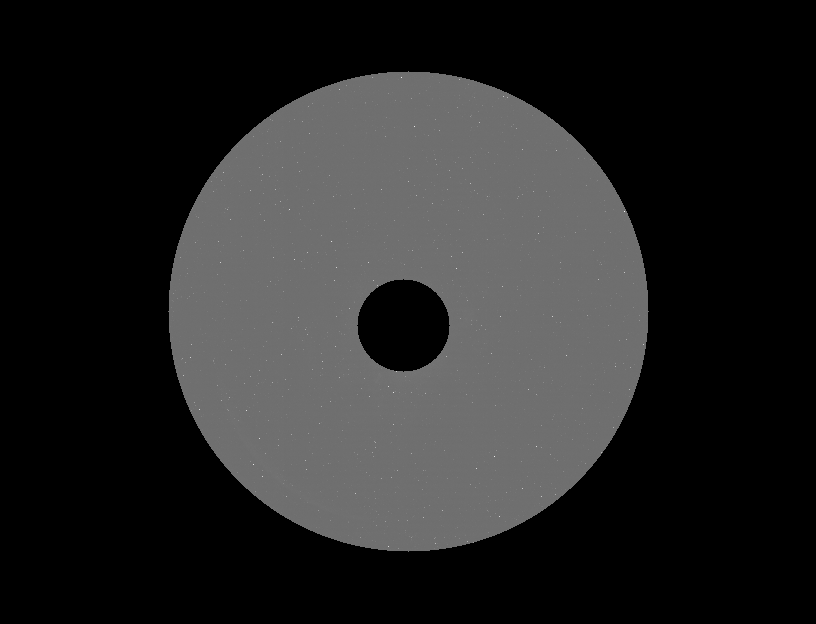}
    \caption{\textbf{Top:} a 10 second calibration image which includes a bright Moon and light reflections from nearby objects. \textbf{Bottom:} Automatically generated bright object mask as well as the user mask which masks areas within the image to prevent false positive detections from occurring}
    \label{fig:radcam-bright-object}
\end{figure}

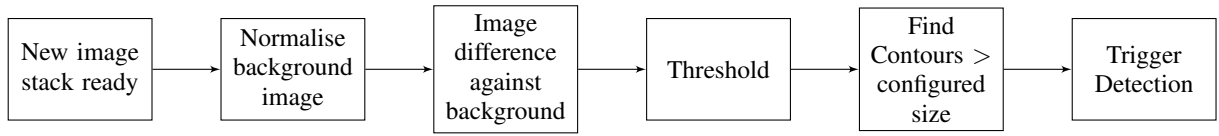
\begin{figure*}[t!] 
\centering
    \begin{tikzpicture}[
        node distance = 2.8cm, 
        auto,]
        \node [block] (box0) {New image stack ready};
        \node [block, right of=box0] (box1) {Normalise background image};
        \node [block, right of=box1] (box2) {Image difference against background};
        \node [block, right of=box2] (box3) {Threshold};
        \node [block, right of=box3] (box4) {Find Contours $>$ configured size};
        \node [block, right of=box4] (box5) {Trigger Detection};

        \path [line] (box0) -- (box1);
        \path [line] (box1) -- (box2);
        \path [line] (box2) -- (box3);
        \path [line] (box3) -- (box4);
        \path [line] (box4) -- (box5);
    \end{tikzpicture}
    \caption{\centering Pipeline for triggering detection}    \label{fig:radcam-detection-pipeline}
\end{figure*}

To detect fireballs, background images are generated by summing a configurable number of images from the camera. Bright object masking based on a configured sigma value of the background image is performed to reduce false positives generated from the Moon. A configurable user mask is also applied to the background image to stop light reflections located outside the image circle of the lens from generating false positives. This mask also prevents foreground moving objects that are close to the horizon from generating false positives. Figure \ref{fig:radcam-bright-object} shows the result of this masking.
In order to reduce the computational resources required to process the high frame rate as well as reduce the prevalence of hot pixels, the detection algorithm is performed on a masked stacked image that consists of a configurable number of summed images that have been masked with the bright object mask. To account for variations in the number of images and their individual exposure and gain settings, the background image is first normalised to the masked stack before the two are differenced. This differenced image is then thresholded by a configurable sigma value and searched for contours based on OpenCV's implementation of \cite{suzuki_topological_1985}. If a contour exists that is larger than a configured size, a detection is triggered. The flow of this detection algorithm is illustrated in the pipeline diagram in Figure \ref{fig:radcam-detection-pipeline}.

\begin{figure}
    \centering
    \includegraphics[width=3cm]{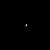} 
    \includegraphics[width=3cm]{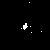} 
    \includegraphics[width=3cm]{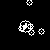} 
    \includegraphics[width=3cm]{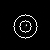} 
    \caption{\textbf{Top left:} Unprocessed image of a fireball. \textbf{Top right:} Thresholded image. \textbf{Bottom left:} Contoured image of which both hot pixels and the fireball have been identified. \textbf{Bottom right:} Aperture selected around the fireball.}
    \label{fig:radcam-aperture-photometry-example}
\end{figure}

To include any images that may have not been bright enough to trigger a detection, a configurable number of images prior to a detection is added to a detection image buffer. In the case of both prototypes, this was set to 1 second worth of images. To reduce the size of the data saved during a detection, these pre-detection images are cropped to 100 x 100 pixels, with the centre position of the crop aligned to the centroid of the current detection contour. All new received images are then added to the detection image buffer after being cropped to 100 x 100 pixels, with the centre position of the crop aligned to the centroid of the current detection contour. If no detection has been triggered for more than 1 second, the detection file is finalised, and the detection state is exited.
Every 10 minutes a 10 second exposure calibration image is acquired, which allows any potential detection to be absolutely calibrated using stars located within the field of view of the image. This calibration image typically reduces useful observation time by less than 1\%.
Initial uncalibrated light curves of the detections are immediately generated using aperture photometry, allowing a user to evaluate whether the event is a fireball without viewing images or videos. Each image within a detection file is thresholded using 3 sigma, the result of which is then searched for contours. The centroid of the largest contour is defined as the position of the fireball in the image, and the minimum enclosing circle diameter is defined as the size of the fireball which is then used to define the aperture size when performing the aperture photometry. Figure \ref{fig:radcam-aperture-photometry-example} shows this process. To produce absolutely calibrated light curves, the detection files are retrieved remotely and manually reduced using an aperture photometry procedure with SkyFit2 \citep{vida_global_2021}.

\subsection{Detection Localised Auto-brightness Control}
To increase the dynamic range of the system beyond the native limitations of the 12-bit sensor, a real-time auto-brightness control algorithm was implemented to adjust the system’s sensitivity during a fireball’s bright flight. Many CMOS cameras include this functionality; however, it is typically based on the average pixel value of the entire frame. This approach is suboptimal for fireball photometry, where the goal is not to correctly expose the night sky, but to ensure a small, transient region of interest maintains a high signal-to-noise ratio (SNR) while avoiding pixel saturation.
The Forrest Airport prototype implements a localised control loop. Once a detection is triggered, the algorithm periodically analyses the detection image, identifying the maximum pixel value within the fireball's contour. This period is user-configurable to balance control responsiveness against computational resources, and was set to 250 ms for the Forrest Airport prototype. Following this analysis, a new gain and exposure value is calculated to adjust the camera's sensitivity, with the goal of driving the measured maximum pixel value towards a user-configurable target value and variance. To define the operational limits of the system, user-configurable minimum and maximum values for both gain and exposure are also set.
For the Forrest Airport prototype system, the available gain ranges from 1 dB to 30 dB, and the exposure period can be varied from a minimum of 20 µs to a maximum of 1800 µs. A consequence of adjusting the exposure period in a fixed frame rate system is a dynamically changing duty cycle (the ratio of exposure time to readout time). At maximum sensitivity (1800 µs exposure), the prototype's duty cycle exceeds 90\%. This decreases to approximately 1\% as the minimum exposure period is approached, though in practice, this would only occur for extremely bright fireballs observed at close range. The control logic is designed to adjust gain before exposure when decreasing sensitivity and increase exposure before gain when increasing sensitivity. The increase in dynamic range of a system using DLAC in astronomical magnitude can be calculated using 

\begin{equation}
    \Delta \text{DR} = 
    2.5 \log_{10}\left(\frac{\text{Exposure}_{\text{max}}}{\text{Exposure}_{\text{min}}}\right) + 
    2.5 \log_{10}\left(\frac{\text{Gain}_{\text{max}}}{\text{Gain}_{\text{min}}}\right)
\end{equation}
where $\Delta \text{DR}$ is the DLAC dynamic range increase in astronomical magnitudes.
The total dynamic range of the system can be approximated by
\begin{equation}
    \text{DR} = 
    2.5 \log_{10}\left(\frac{\text{FWC}}{\sigma_{\text{read}}}\right) + 
    \Delta \text{DR}
\end{equation}
where $\text{FWC}$ is the Full Well Capacity (e$^-$), $\sigma_{\text{read}}$ is the Read Noise (e$^-$), and $\Delta \text{DR}$ is the DLAC dynamic range increase defined above.
In the case of the Forrest Airport prototype, DLAC has increased the dynamic range of the system by 8.51 magnitudes. The full well capacity of the Allied Vision Alvium U-052 with a gain of 30dB is 3162e-, and the read noise is approximately 21.8e-, yielding an initial dynamic range of 5.4 magnitudes and a final dynamic range of approximately 13.91 astronomical magnitudes. It should be noted that the read noise parameter used here is a nominal value and may vary based on factors such as ambient temperature conditions; however, this calculation provides a reasonable approximation of the system's total dynamic range. This increased dynamic range would allow a single camera to capture the entire bright flight of most fireballs. However, due to the rapid brightness changes inherent to fireballs, it can be challenging for such a system to react instantaneously, and some pixel saturation may still occur during abrupt flaring events.
\section{System Characterisation and Performance}
\subsection{Photometric Validation using the Moon as a Standard}

\begin{figure*}
    \centering
    \includegraphics[width=\textwidth]{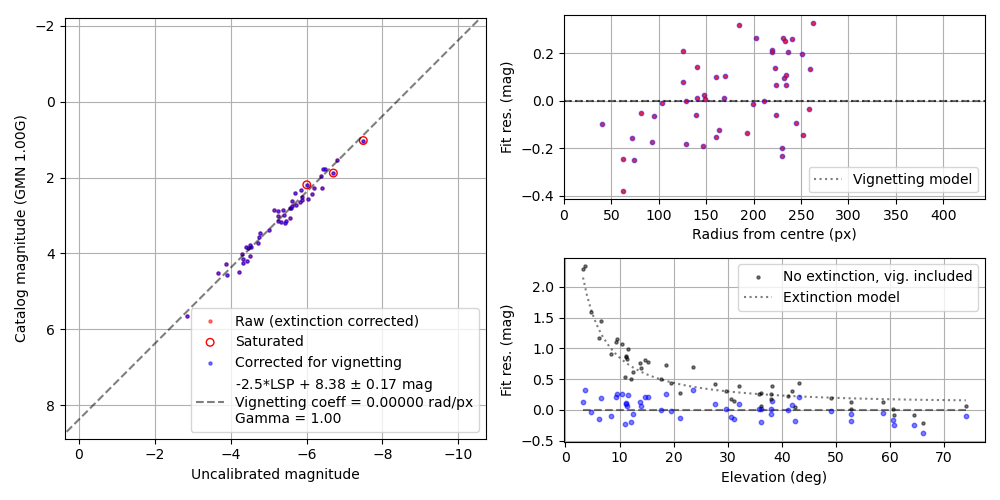}
    \caption{\centering Photometric fit of the long exposure calibration image from the Forrest prototype taken on at 2025-09-30 at 12:59:44. LSP represents the logarithm of the summed pixel intensity within the aperture used to estimate brightness.}
    \label{fig:moon-photometric-fit}
\end{figure*}

\begin{figure}
    \centering
    \includegraphics[width=\columnwidth]{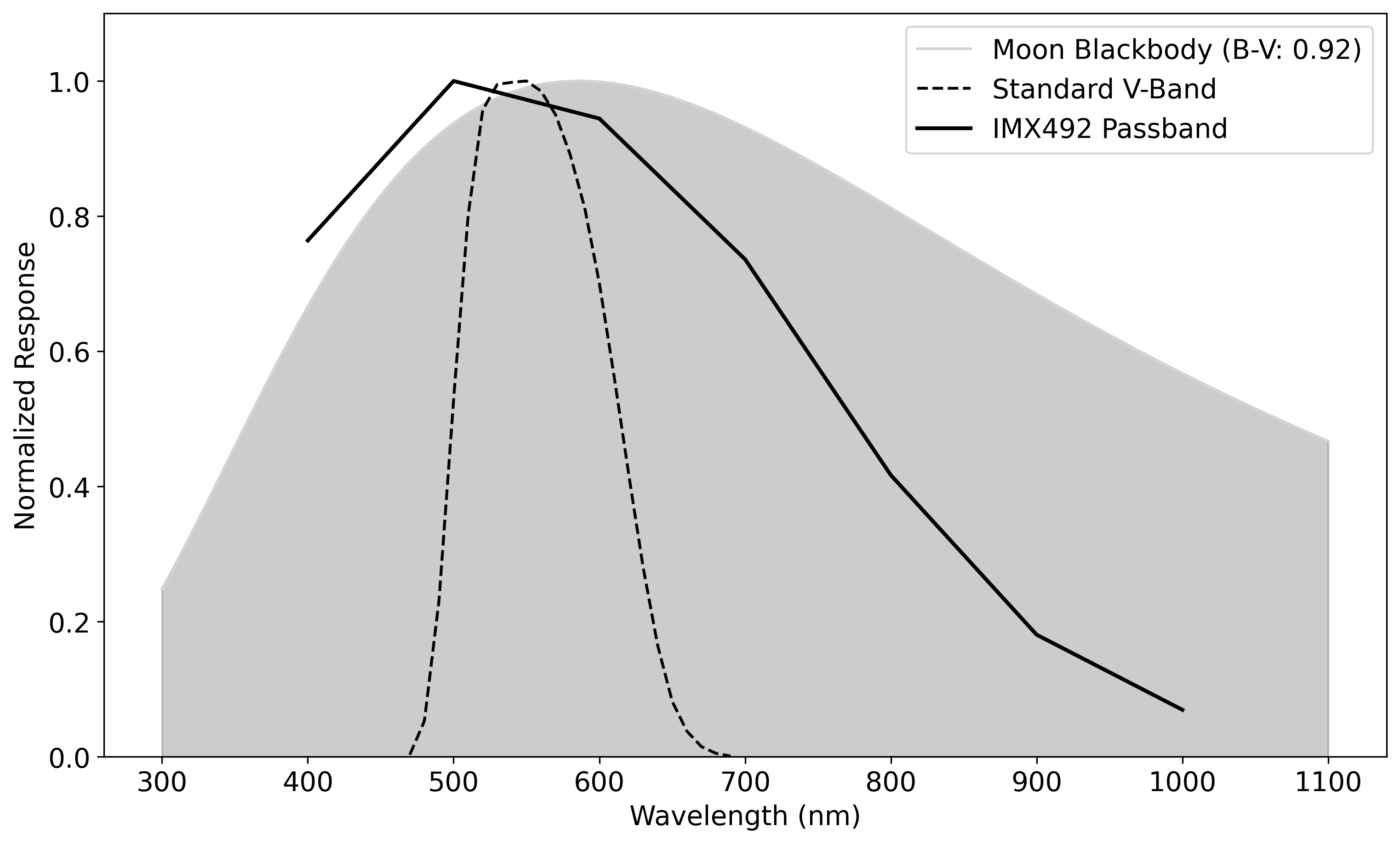}
    \caption{V-band passband and IMX426 passband used to calculate the colour correction for the NASA JPL Horizons V-band magnitude of the Moon as observed from Forrest Airport at 2025-09-30 13:02:28 UTC}
    \label{fig:moon-passband-response}
\end{figure}

Validating the photometric accuracy of the system at the very short exposure times used during fireball events is challenging, as these brief frames do not capture enough background stars for a direct calibration. To avoid this, an indirect validation was performed using the Moon as a stable, bright calibration source.
The procedure involved capturing two distinct images with the Forrest prototype. The first was a 10-second calibration exposure taken at 20 dB gain on 2025-09-30, ending at 12:59:44 UTC to ensure a sufficient starfield was captured. Figure \ref{fig:moon-photometric-fit} shows the photometric fit generated using SkyFit2, which yielded a systematic error of ±0.17 magnitudes and an atmospheric extinction coefficient of 0.6. 
A second, short-exposure image was then taken on 2025-09-30 at 13:02:28 UTC (1800 µs exposure, 0 dB gain). This configuration was chosen to ensure the Moon was brightly illuminated but not saturated. Aperture photometry was performed on the Moon in this second image. Using the Forrest prototype's location and the precise time of the image, the Moon's altitude, azimuth, and distance from the observer were calculated using Astropy \citep{robitaille_astropy_2013}. This data, combined with the previously determined extinction coefficient of 0.6, was used to correct the measurement for extinction. The resulting apparent magnitude of the Moon in the IMX426 passband was calculated to be -11.23.
To verify this measurement, an independent reference magnitude was obtained from the NASA JPL Horizons system for the same time and observer location, which gave an extinction corrected apparent V-band magnitude of -10.289. A colour correction was then calculated to transform this V-band magnitude to the specific spectral response passband of the Sony IMX426 sensor assuming a Moon B-V index of 0.92. Figure \ref{fig:moon-passband-response} shows the relative spectral responses used in this conversion. The resulting colour-corrected IMX426-passband reference magnitude for the Moon was -11.195.
Table \ref{tab:moon-calibration} shows a summary of the results. The difference between our system's measured apparent magnitude (-11.23) and the colour-corrected JPL Horizons reference value (-11.195) is just 0.035 magnitudes. This deviation is well within the ±0.17 magnitude systematic error of the SkyFit2 photometric fit, thus validating the photometric accuracy of the instrument at the short exposure settings used for fireball observations.
\begin{table}
\caption{Moon Photometric Calibration Results}
\centering
\begin{tabular}{rrrrrrrrrr}
\hline
 Moon B-V Index & 0.92 \\
 Moon V magnitude (JPL Horizons) & -10.289 \\
 Color Correction (IMX426 Mag - V Mag) & -0.986 \\
 Moon IMX426 magnitude (JPL Horizons) & -11.195 \\
 Moon IMX426 magnitude (SkyFit2) & -11.23 \\
 \hline
 Magnitude error & 0.035 \\
\hline
\end{tabular}
\label{tab:moon-calibration}
\end{table}

\subsection{Photometric Validation and Comparison with a 30 FPS System}
\begin{figure}
    \centering    \includegraphics[width=\columnwidth]{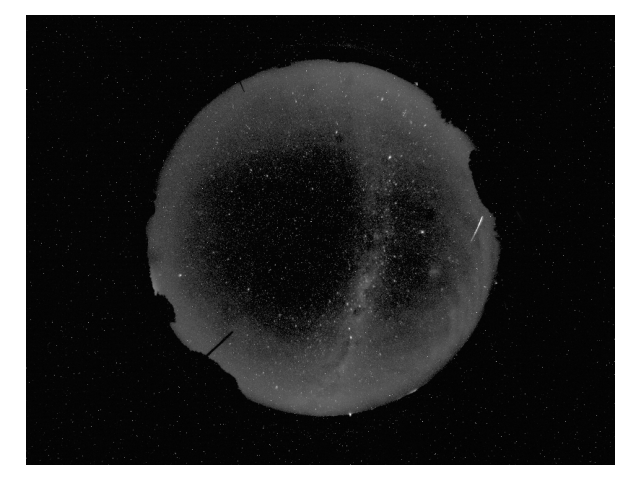} 
    \caption{The fireball observed on May 6, 2024. The image is the sum of all the detection frames composited over the log stretched calibration image.}
    \label{fig:radcam-fireball-2024-05-06}
\end{figure}

\begin{figure}
    \centering
    \includegraphics[width=\columnwidth]{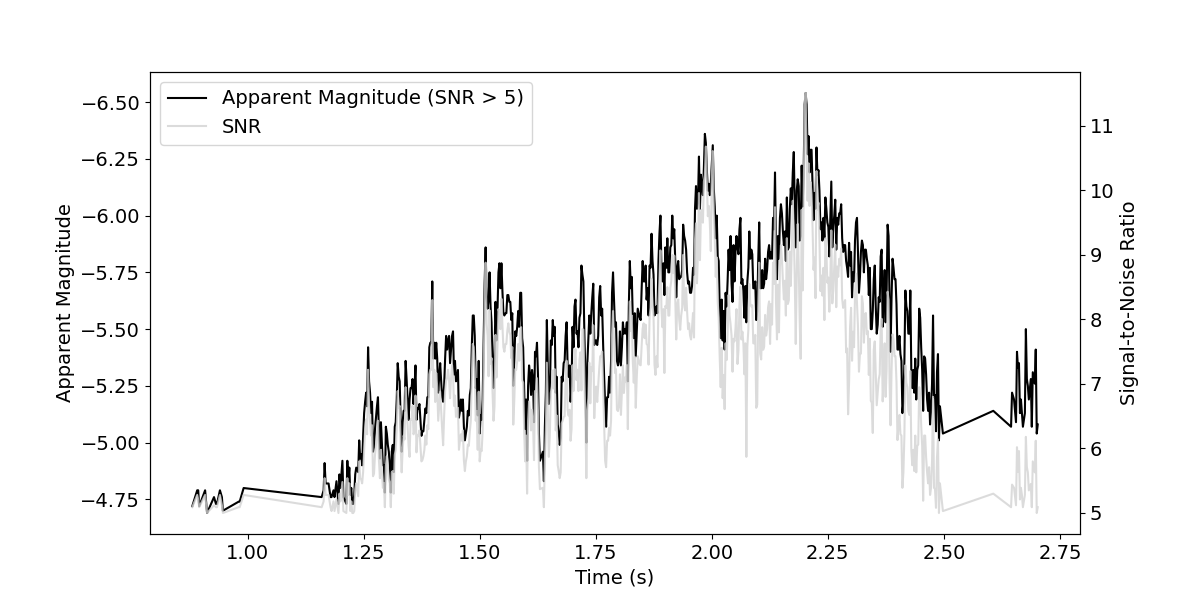}
    \caption{20240506 SNR clipped extinction corrected apparent magnitude from the prototype.}
    \label{fig:radcam-fireball-2024-05-06-apparent}
\end{figure}

\begin{figure*}[ht!]
    \centering
\includegraphics[width=\textwidth]{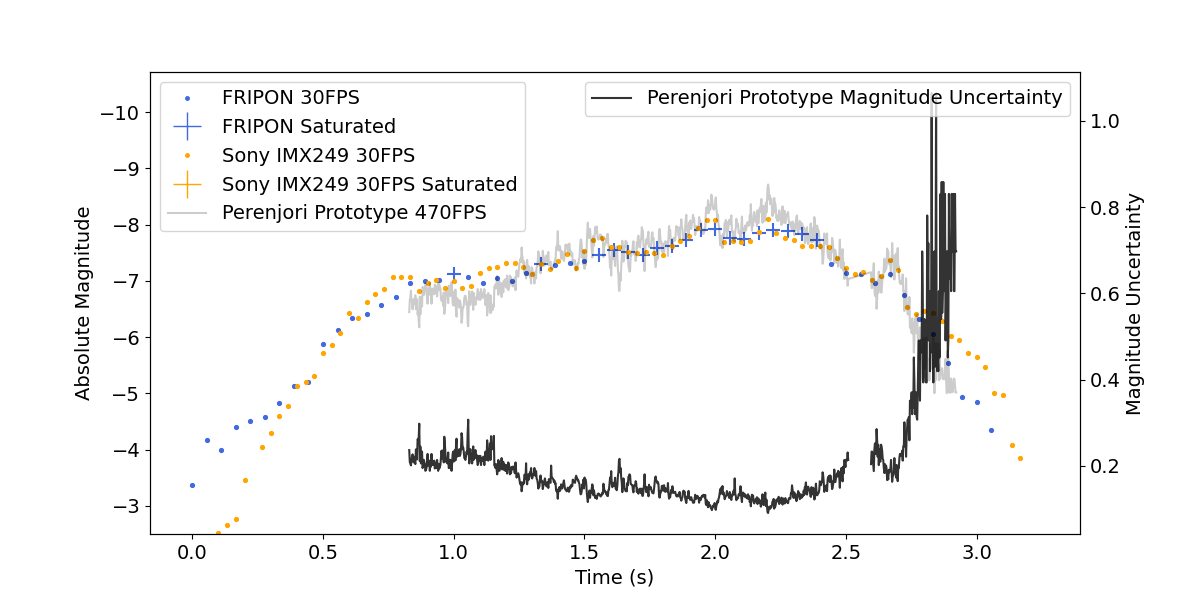}
    \caption{20240506 absolute light curves from the Perenjori prototype (Sony IMX426), the co-located Sony IMX249 camera, and the FRIPON camera (Sony  ICX445ALA).}
    \label{fig:radcam-fireball-2024-05-06-light-cruve2}
\end{figure*}

\begin{figure*}[ht!]
    \centering
    \includegraphics[width=\textwidth, trim=0 0 0 0.3cm, clip]{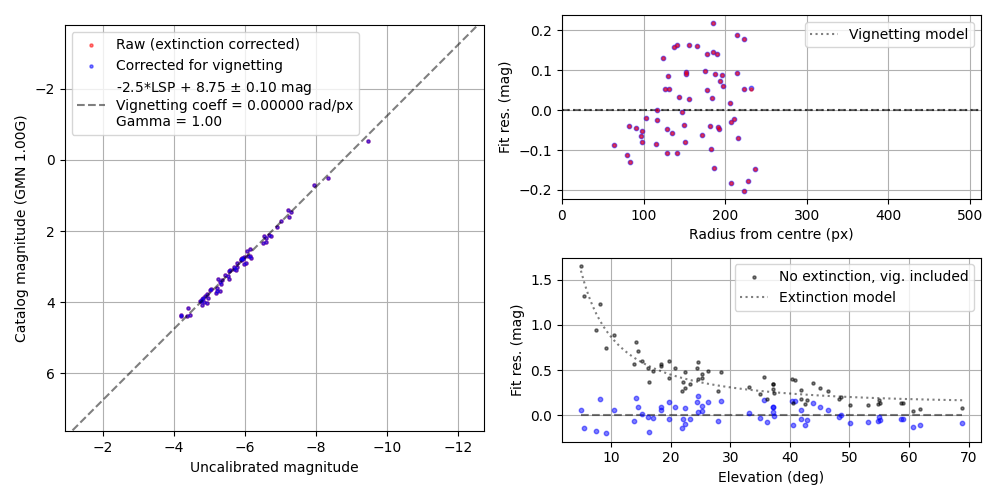} 
    \caption{\centering20240506 photometric fit of the calibration image from the Perenjori prototype.}
    \label{fig:radcam-fireball-2024-05-06-photometric-fit}
\end{figure*}

On May 6, 2024, at 12:24 UTC, a fireball was detected west of Perth by the Perenjori prototype as shown in Figure \ref{fig:radcam-fireball-2024-05-06}. The event was simultaneously observed by the co-located 30 FPS Sony IMX249 camera, a nearby FRIPON station \citep{colas_fripon_2020}, and several Global Meteor Network (GMN) cameras \citep{vida_global_2021}. This multi-instrument observation provides an opportunity to validate the photometric performance and accuracy of the high frame rate system by directly comparing its derived light curve against those from established, lower-rate systems.
 The trajectory of this fireball was calculated using GMN camera data and is shown in Figure \ref{fig:stations}. Figure \ref{fig:radcam-fireball-2024-05-06-apparent} shows the apparent magnitude corrected for extinction produced using data of the event where data points with an SNR of less than 5 have been removed. The SNR has been calculated according to \citep{howell_two-dimensional_1989}. A software defect in an early version of the software resulted in gaps at the beginning and middle of the light curve. A minimum sensitivity for the event of apparent magnitude -4.7 can be observed. Figure \ref{fig:radcam-fireball-2024-05-06-light-cruve2} shows the absolutely calibrated light curves from the Perenjori prototype (Sony IMX426), the co-located Sony IMX249 camera, and the FRIPON camera (Sony ICX445ALA), as manually reduced using SkyFit2. For this specific event, data recording anomalies including dropped frames in the FRIPON and IMX249 video feeds, as well as a file segmentation issue that split the prototype’s detection into separate files, which prevented the use of absolute timestamps. Consequently, the light curves have been manually time-aligned. While this results in a visible temporal offset, this event represents the only simultaneous capture by all three instruments and is included here to validate the system's photometric accuracy rather than its absolute timing capabilities. Data points that had apertures containing saturated pixels have been marked to show where intensity values have been underestimated. With a peak absolute magnitude of around -8, the high-resolution light curve demonstrates acceptable SNR characteristics with a random magnitude uncertainty of less than 0.3 for the majority of the bright flight, though this uncertainty degrades to greater than 1 magnitude in the final 0.5 seconds. The SkyFit2 photometric fit can be found in Figure \ref{fig:radcam-fireball-2024-05-06-photometric-fit} and shows a systematic magnitude uncertainty of $\pm$0.10. The absolute light curves show agreement within $\pm$0.5 magnitude in the unsaturated regions of all cameras. Discrepancies between the light curves could be attributed to random uncertainty associated with the SNR of the high frame rate camera, influenced by the relatively dim fireball, as well as differences in spectral sensitivity between the sensors.

\subsection{Application of High-Cadence Data to Fragmentation Analysis}
To demonstrate the scientific application of the high-cadence data, a bright fireball captured by the Forrest prototype is presented. This system is equipped with the DLAC algorithm, designed to capture the full dynamic range of a fireball's bright flight without saturation. This combination of high temporal resolution and active brightness management yields a complete, unsaturated light curve for most of the fireball's bright flight, providing the essential data for both detailed fragmentation analysis and advanced ablation modelling from a single instrument.

\begin{figure*}[ht!]
    \centering
    \includegraphics[width=\textwidth]{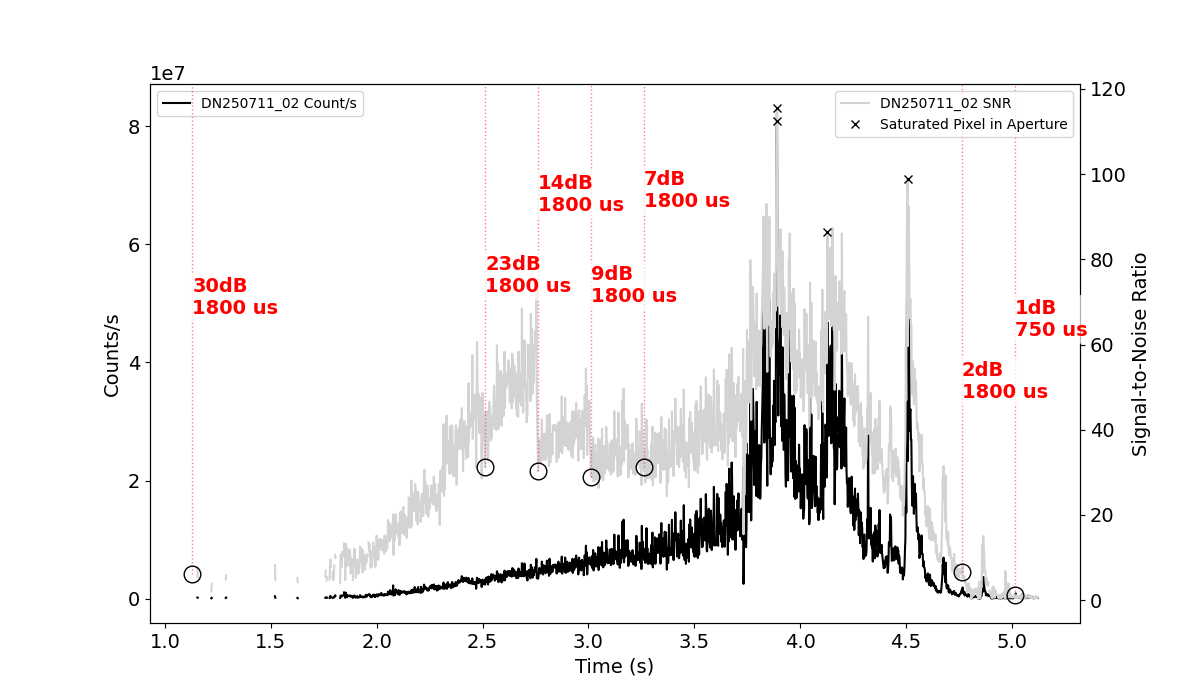} 
    \caption{\centering DN250711\_02 uncalibrated light curve, SNR, and DLAC parameters from the Forrest prototype}
    \label{fig:radcam-fireball-2025-07-11-uncalibrated-lightcurve}
\end{figure*}

\begin{figure}
    \centering
    \includegraphics[width=\columnwidth]{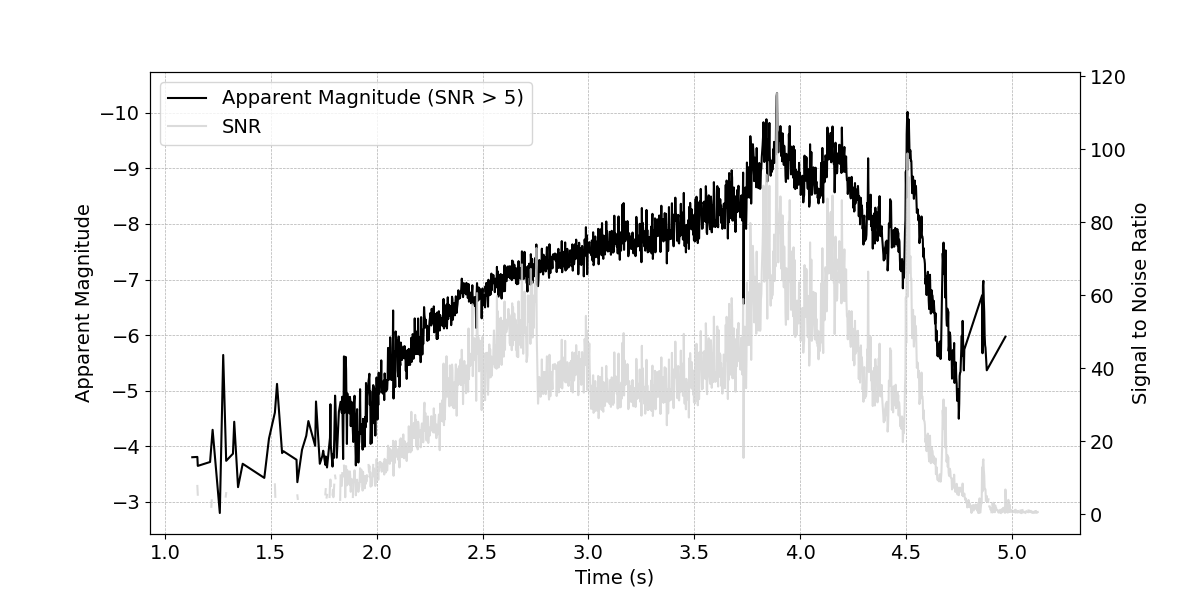} 
    \caption{DN250711\_02 SNR clipped extinction corrected apparent magnitude from the Forrest prototype}
    \label{fig:radcam-fireball-2025-07-11-limiting}
\end{figure}

\begin{figure*}[ht!]
    \centering
\includegraphics[width=\textwidth]{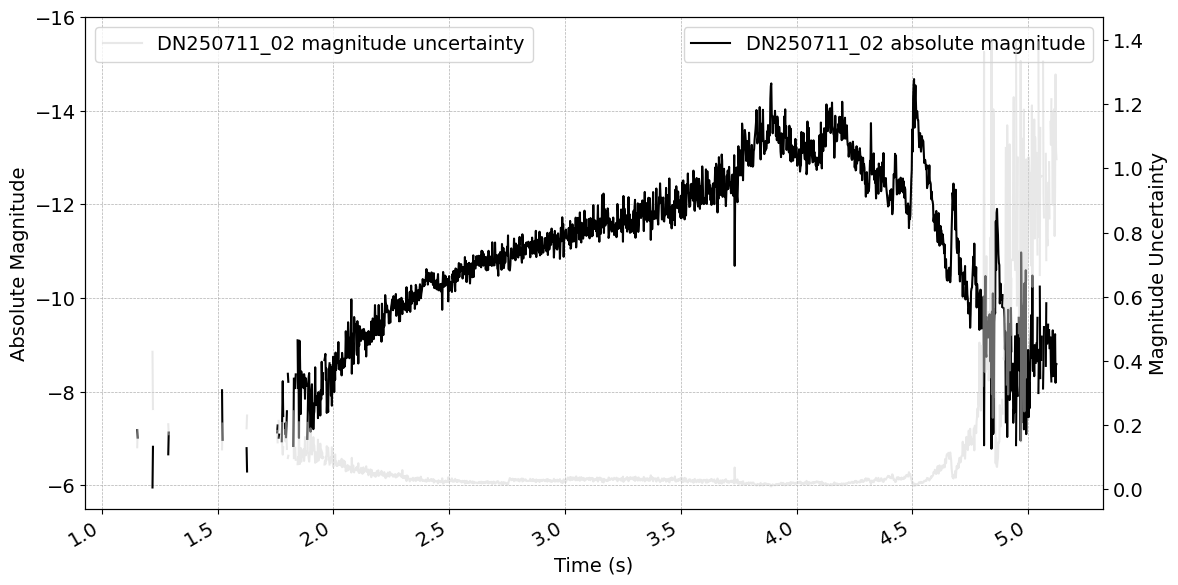}
    \caption{DN250711\_02 absolute light curve with magnitude uncertainty from the Forrest prototype. Magnitude uncertainty for unseen areas is typically less than 0.1.}
    \label{fig:radcam-fireball-2025-07-11-absolute}
\end{figure*}

\begin{figure*}[ht!]
    \centering
\includegraphics[width=\textwidth]{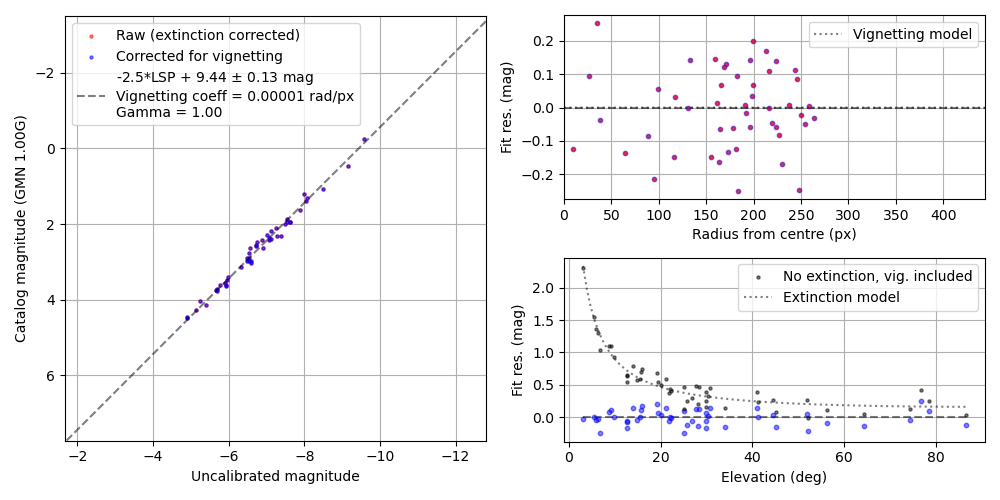} 
    \caption{DN250711\_02 photometric fit of the calibration image from the prototype. LSP represents the logarithm of the summed pixel intensity within the aperture used to estimate brightness.}
    \label{fig:radcam-fireball-2025-07-11-photometric-fit}
\end{figure*}

On July 11, 2025 at 13:03UTC a fireball (DN250711\_02) was detected northwest of Forrest Airport by the second prototype, as well as several DFN stations. 
The trajectory was calculated using DFN station data and is shown in Figure \ref{fig:stations}. 
The uncalibrated light curve and SNR are shown in Figure \ref{fig:radcam-fireball-2025-07-11-uncalibrated-lightcurve}. 
Saturated pixels are marked as well as changes to the gain and exposure due to the implementation of DLAC. Increases in gain and exposure period after the peak brightness of the fireball did not occur because of an earlier implementation. The starting gain and exposure was 30 dB and 1800 µs respectively, while the final gain and exposure was 1dB and 750 µs. With these changes, the dynamic range of the camera was increased by approximately 4.5 astronomical magnitudes during the fireball's bright flight, and minimised saturation of the sensor to 4 discrete data points. A minimum sensitivity of apparent magnitude -2.8 and a maximum apparent magnitude of -10.36 can be observed in the SNR clipped light curve shown in Figure \ref{fig:radcam-fireball-2025-07-11-limiting}. Figure \ref{fig:radcam-fireball-2025-07-11-absolute} shows the absolutely calibrated light curve from the prototype, with a peak magnitude of nearly -15. The photometric fit from SkyFit2 can be found in Figure \ref{fig:radcam-fireball-2025-07-11-photometric-fit} and shows a systematic magnitude uncertainty of $\pm$0.13. 
The fireball fragmentation modelling is based on the observed light curve and the fireball dynamics. It has been performed using our implementation \citep{vida_direct_2023, vida_first_2024} of the \citet{borovicka_kosice_2013} semi-empirical fragmentation model and has been applied effectively to many meteors of various compositions \citep{borovickaInstrumentallyRecordedFall2015, brown_golden_2023, mcmullanWinchcombeFireballThat2024}. In this model, meteoroid ablation is assumed to proceed mainly through fragmentation caused by the release of discrete fragments, either via the splitting of the main body or the ejection of mm-sized and smaller dust. Each ejected fragment is numerically integrated using the classical single-body ablation equations.

Table \ref{tab:model_phys} summarises the best-fit global physical and dynamical parameters of the meteoroid based on the model comparison to the data. In addition, Table \ref{tab:fragmentation} presents a version of the modelled fragmentation behaviour needed to explain the features in the light curve. We emphasise that this is not a unique solution but simply representative. The general fitting procedure we adopted is described in more detail in \citet{vida_first_2024} where it has been successfully applied. The best fit was achieved by assuming a bulk density appropriate to a carbonaceous chondrite.

\begin{figure}
\centering
\includegraphics[width=\columnwidth]{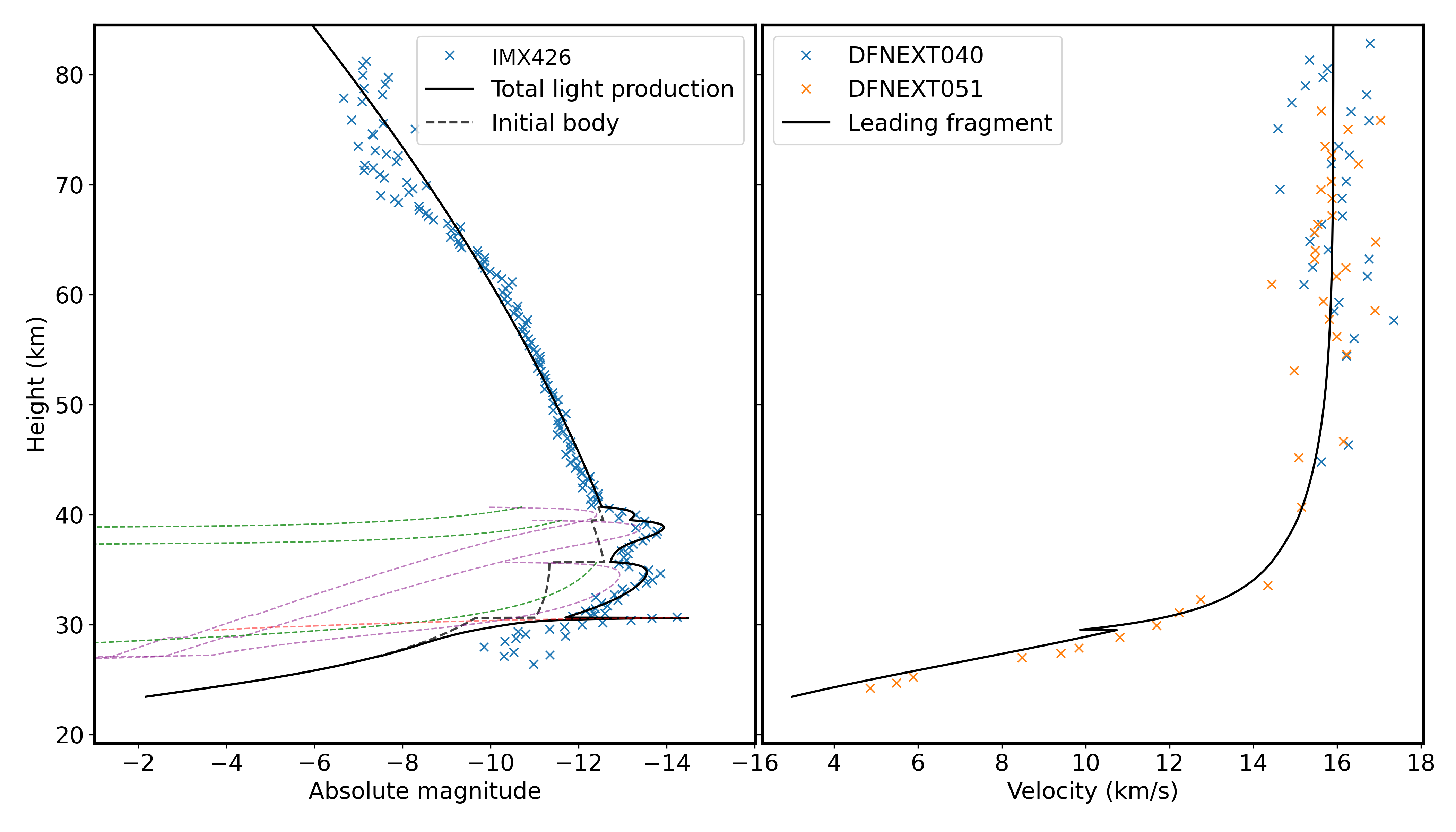}
\caption{Left: Measured DN250711\_02 fireball light curve as a function of height from instrumental observations (blue crosses) as compared to the total light production estimated from the semi-empirical model (solid black line). The individual light curves for eroding fragments (green dashed line) and dust released from eroding fragments (purple dashed line) are also shown. The modelled light production from the ablation of the main mass is given by the dashed black line. Right: The measured point-to-point velocities for stations DFNEXT040 and DFNEXT051 as compared to the model estimate of the velocity for the leading fragment as a function of height.}
\label{fig:simulation_comp}
\end{figure}

Figure \ref{fig:simulation_comp} shows the comparison between the model and the observations. For the purposes of best comparisons with the model, the light curve has been averaged to a cadence of 0.02 seconds, matching the time step of the model. The model optimally fits the dynamics and successfully reproduces the main features of the light curve. The first fragmentation event is visible in the light curve at 70 km but has not been modelled. We attempted to include it in the model fit, but that required lowering the initial mass, which made it insufficient to reach the terminal height and explain the rest of the light curve. We suspect that the fit mismatch at the beginning is caused by the effect of preheating, which has been observed in other fireballs such as Žďár nad Sázavou and 2023 CX1 \citep{spurny_zdar_2020, egal_catastrophic_2025}.
\begin{table}
\caption{Model-inferred physical and dynamical properties of the 2025-07-11 fireball. A grain density of 2700 kg~m$^{-3}$ was used in the model.}
\centering
\begin{tabular}{llrl}
\hline
Description & & Value\\
\hline\hline 
Initial mass (kg)                                   & $m_0$      & 80.0   \\
Initial speed at 180~km (km~s$^{-1}$)               & $v_0$      & 15.91  \\
Geocentric zenith angle                             & $Z_c$      & $7.09^{\circ}$ \\
Bulk density (kg~m$^{-3}$)                          & $\rho$     & 2200   \\
Grain density (kg~m$^{-3}$)                         & $\rho_g$   & 2700   \\
Shape-density coefficient                           & $\Gamma A$ & 1.21   \\
\hline
\end{tabular}
\label{tab:model_phys}
\end{table}

\begin{figure}
\centering
\includegraphics[width=\columnwidth]{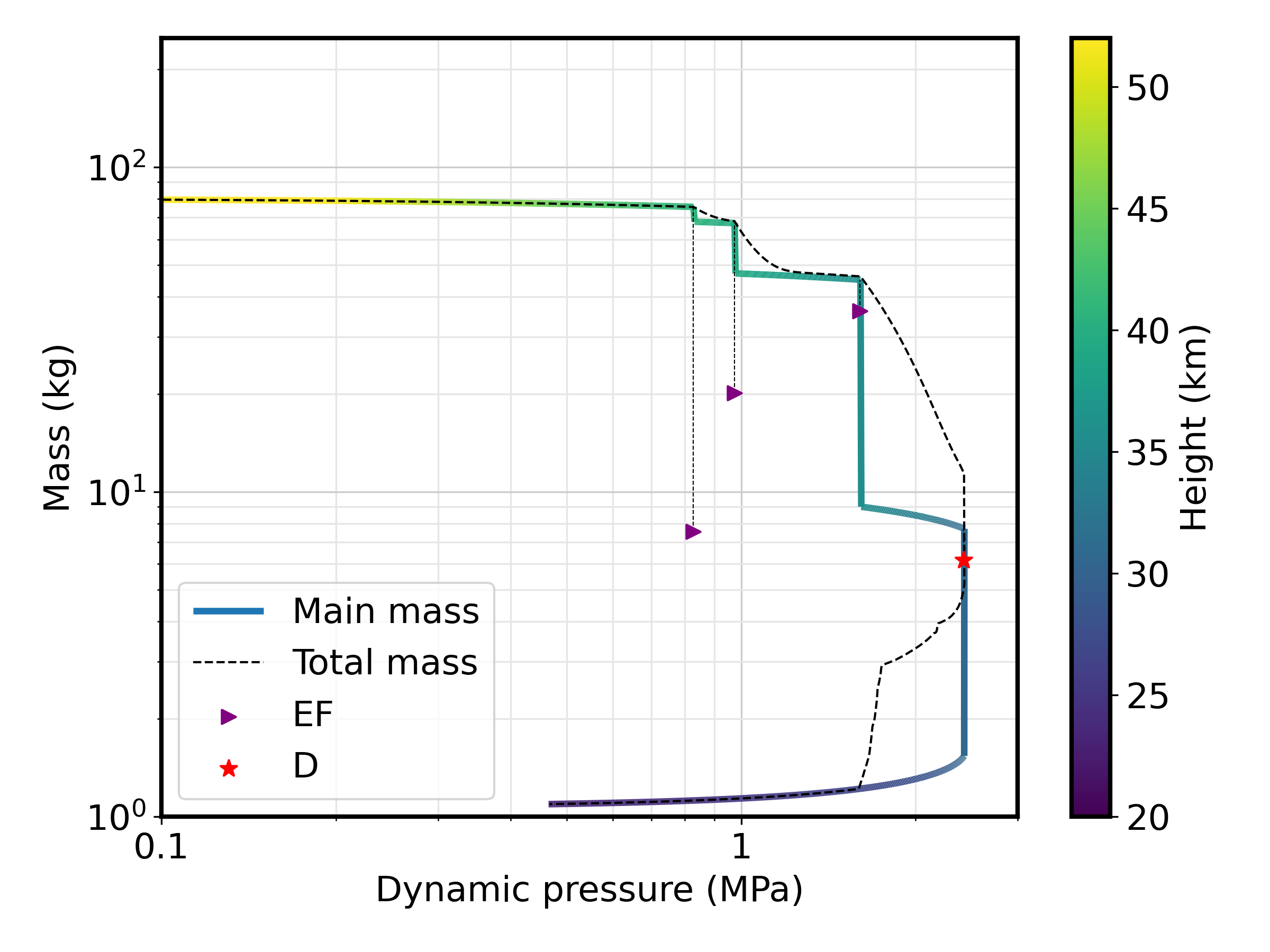}
\caption{The amount of mass remaining in the main fragment as a function of dynamic pressure. This shows the mass loss by fragmentation mode, either leading to a release of an eroding fragment (EF) or dust (D). The fireball height is colour-coded. A drag coefficient of $\Gamma = 1.0$ was used to compute the dynamic pressure for consistency with previous work \citep{borovicka_two_2020}.} 
\label{fig:fragmentation}
\end{figure}

The fragmentation behaviour of the fireball is characterised by three eroding fragments (EF) events followed by a single disruptive dust (D) release. The eroding fragments, released between 41 and 36 km altitude, produce a gradual brightening, while the final disruption at 30.7~km is responsible for the terminal flare, which is explained by the almost complete crumbling of the meteoroid (80\% of the remaining mass) into mm-sized dust at a dynamic pressure of ~2 MPa.
Figure \ref{fig:fragmentation} shows the modelled mass loss as a function of dynamic pressure in the main fragment. The fragmentation begins at a dynamic pressure of 0.82~MPa, with more significant mass loss events occurring above 0.9~MPa. This is broadly consistent with the results of \cite{borovicka_two_2020} for fireballs estimated to have produced ordinary chondrites where severe fragmentation typically occurred at dynamic pressures of $\gtrsim0.9$~MPa, although we had to assume a carbonaceous composition to fit the brightness and the total amount of deceleration. The final disruptive event occurs at a peak dynamic pressure of 2.08~MPa. The final modelled mass of the fragment which survived the atmospheric flight is 1.54~kg.

\begin{table*}
\caption{Modelled fragmentation behaviour. The fragment mass percentage in the table is referenced to the mass of the main fragment at the moment of ejection. The mass distribution index for all grains was $s = 2.0$ (see a discussion in \citealt{vida_first_2024} for how this parameter affects the fit). The values of the dynamic pressure are computed using a drag coefficient of $\Gamma = 1.0$.}
\centering
\begin{tabular}{rrrrrrrrrr}
\hline
Time$\mathrm{^{a}}$  & Height & Velocity             &  Dyn pres & Main $m$ & Fragment  & $m$ & $m$     & Erosion coeff         & Grain $m$  \\
(s)   & (km)   & ($\mathrm{km~s^{-1}}$) &  (MPa)    & (kg)        &           & (\%) & (kg)    & ($\mathrm{kg~MJ^{-1}}$) & range (kg) \\
\hline\hline 
 2.80 & 40.70 & 15.17 & 0.821 & 75.51 & EF & 10.0  & 7.551 & 0.50 & $10^{-4} - 10^{-2}$ \\
 2.88 & 39.50 & 15.04 & 0.940 & 67.33 & EF & 30.0  & 20.199 & 0.50 & $10^{-4} - 10^{-2}$ \\
 3.14 & 35.70 & 14.44 & 1.431 & 45.06 & EF & 80.0  & 36.051 & 0.10 & $10^{-4} - 10^{-2}$ \\
 3.51 & 30.70 & 12.07 & 2.075 &  7.70 & D  & 80.0  & 6.163 & -    & $10^{-6} - 10^{-4}$ \\
\hline
\multicolumn{10}{l}{$\mathrm{^{a}}$ Seconds after 2025-07-11 13:03:09.812475 UTC.} \\
\multicolumn{10}{l}{EF = New eroding fragment; D = Dust ejection.} \\
\end{tabular}
\label{tab:fragmentation}
\end{table*}

\section{Discussion}
Video camera systems typically use global shutter machine vision cameras with no moving parts, contrasting them to long exposure methods that require either a rotating mechanical assembly, or the insertion of an electro-optical element between the lens and sensor in addition to the associated driving electronics \cite{howie_submillisecond_2017}. PMT sensors require high voltage power supplies to operate, while the Allied Vision Alvium U-052 is powered directly from the same USB3.0 port that is used for data transfer. 
The use of a short exposure period enabled by the high frame rate allows the camera to image brighter fireballs without saturating. The high frame rate increases the temporal resolution of the resulting light curve, allowing the identification of bright flare events associated with fragmentation similar to PMT sensors. Video camera systems typically require more computational resources to detect and process imagery compared to long exposure camera and all-sky radiometers due to the larger data throughput of the system, however, the detection software used for the prototype was found to use less than 10\% of a single CPU core on a 12th generation Intel i7 12700K processor produced in 2021 which has a base CPU frequency of 3.6 GHz. 
The most significant challenge for fireball photometry using video imaging is the limited dynamic range of CMOS sensors, which often leads to saturation during the brightest phases of an event. To address this, the Forrest prototype successfully implemented the DLAC algorithm. This system was shown to prevent saturation by dynamically adjusting the camera's exposure and gain in real-time, based on the brightness of the fireball within its detection aperture. This active management of sensitivity largely solves the dynamic range issue, extending the dynamic range of the system by 8.5 astronomical magnitudes, giving a total dynamic range of 13.91. This effectively provides a single, low-cost imaging sensor with a total dynamic range comparable to that of dedicated photomultiplier tubes, while retaining the ability to perform its own astrometric and photometric calibration.
The analysis of the May 6, 2024 fireball allows for a direct comparison between the high frame rate prototype and established 30 FPS systems, including a co-located camera and a nearby FRIPON station. The results demonstrate that all systems are capable of producing accurately calibrated absolute light curves, with photometric agreement within ±0.5 magnitudes in their unsaturated regions. This validates the fundamental performance of the high frame rate approach for calculating the photometric mass of fireballs. However, while this photometric mass can be generated from all systems, the data required for advanced scientific analysis, such as semi-empirical fragmentation modelling, necessitate the higher temporal resolution that this system provides. This level of modelling depends on resolving the fine-scale details of flare events and rapid brightness changes that characterise fragmentation.
The detailed fragmentation modelling of the July 11, 2025 fireball, made possible by the high-cadence light curve, revealed the meteoroid's progressive disintegration through the atmosphere. The model indicates the 80 kg body began fragmenting under a dynamic pressure of 0.82 MPa, followed by more significant mass loss events. The simulation successfully reproduced the light curve's primary features, attributing the terminal flare to a major disruption at 2.08 MPa, where 80\% of the remaining mass crumbled into dust. However, while the model captures the overall structure, it currently lacks the fidelity to explain all the fine-scale variations observed in the high-frequency light curve; the sharpness of the individual peaks is particularly important as it provides constraints on the size distribution of the grains released during fragmentation events. This limitation arises partly because the current implementation of the fragmentation model operates at a maximum time step of 0.02 s (50 Hz), requiring the ~500 Hz instrument data to be averaged, which smooths out the finest temporal details. Interestingly, while these fragmentation pressures are broadly consistent with stronger ordinary chondrites, the model required a lower bulk density, typical of a carbonaceous chondrite, to simultaneously fit the object's brightness and deceleration. This demonstrates that the high temporal resolution data is crucial for constraining the physical properties and fragmentation behaviour, providing insights that would be unattainable with lower-rate systems.
Beyond identifying discrete fragmentation events, high temporal resolution light curves have also revealed flickering, and quasi-periodic brightness variations, which have been attributed to mechanisms such as the rotation of irregular meteoroids and self-induced fluctuations in ablation \citep{spurnyEN310800VimperkFireball2001, beech_meteoroid_2001, mancusoMeteoroidRotationQuasiperiodic2024, babadzhanov_features_2004}. These recurrent modulations offer further insight into the shape, spin state, and physical stability of meteoroids as they ablate. Short, millisecond-duration intensity spikes have been observed in high-frequency radiometric records and are believed to be the result of triboelectric charging effects during atmospheric entry \citep{spurnyElectricChargeSeparation2008}. While standard 30 FPS video observations cannot resolve these features, high frame rate imaging significantly improves temporal resolution, bridging the gap between standard video and PMT-based radiometers. This capability allows for the investigation of rapid brightness fluctuations that were previously inaccessible to imaging systems. By deploying these instruments at DFN stations, the resulting large dataset will combine high-cadence light curves with the network's highly accurate trajectory and dynamics data, enabling a more detailed investigation into the physical mechanisms behind these high-frequency fluctuations.

\section{Conclusion}
High temporal resolution photometry, enabled by high frame rate cameras featuring DLAC, provides a complete and saturation-minimised view of the dynamic behaviour of fireballs, particularly during fragmentation and flare events. The instrument's photometric accuracy was validated using the Moon as a bright calibration source. Our measurement agreed to within 0.035 magnitudes of the colour-corrected reference value, confirming the system's accuracy at the short exposure settings used for fireball observations. The absolute light curve from the May 6, 2024 event agrees within ±0.5 magnitude when compared to co-located systems, while resolving short-duration flares not captured by conventional 30 FPS cameras. Furthermore, the successful implementation of the DLAC algorithm was demonstrated with the July 11, 2025 event, where a bright fireball with a peak absolute magnitude of nearly -15 was recorded with minimal saturation, extending the system's observable dynamic range to an approximate total of 13.9 magnitudes. The resulting high-quality light curve provided the necessary data to perform semi-empirical fragmentation modelling, demonstrating the system's capability as a standalone instrument for detailed physical analysis. Future work will include the integration of direct Global Navigation Satellite System time synchronisation for sub-millisecond absolute timing, further refinement of the DLAC algorithm to minimise response times, the evaluation of the system for daytime fireball detection, and the deployment of this system in a hybrid configuration alongside dedicated high-resolution astrometric cameras to create optimal next-generation GFO instrumentation.

\subsubsection*{Acknowledgements}
We thank Peter McKellar for providing the fireball trajectory data used in this paper. The authors used Gemini to improve the readability and language of this manuscript. The authors reviewed and revised the output and take full responsibility for the content.

\subsubsection*{Data Availability Statement}
Supplementary materials have been uploaded as a Zenodo record at \href{https://doi.org/10.5281/zenodo.15347347}{https://doi.org/10.5281/zenodo.15347347}.

\bibliographystyle{paslike}
\bibliography{references}{}

\end{document}